\documentstyle[twocolumn,aps,prb,epsf]{revtex}
\begin{document}
\newcommand{\beq}{\begin{equation}}
\newcommand{\eeq}{\end{equation}}

\title{AC resistivity of $d$-wave ceramic superconductors}

\author{Mai Suan Li$^{1,2}$ and Daniel Dom\'{\i}nguez$^3$}

\address{$^1$Institute of Physics, Polish Academy of Sciences,
Al. Lotnikow 32/46, 02-668 Warsaw, Poland\\
$^2$ Institut f{\H u}r Theoretische Physik, Universit{\H a}t zu K{\H o}ln,
Z{\H u}lpicher Stra{\ss}e 77, D-50937 K{\H o}ln, Germany\\
$^3$ Centro At\'{o}mico Bariloche, 8400 S. C. de Bariloche, Rio Negro,
Argentina}

\address{
\centering{
\medskip\em
{}~\\
\begin{minipage}{14cm}
We model  $d$-wave ceramic superconductors 
with a three-dimensional lattice of randomly distributed
$\pi$ Josephson junctions with finite self-inductance. 
The linear and nonlinear ac resistivity of the 
$d$-wave ceramic superconductors is obtained as function of temperature
by solving the corresponding Langevin dynamical equations. 
We find that the linear ac resistivity  remains finite at the
temperature $T_p$ where the third harmonics of resistivity has a peak. 
The current amplitude dependence of the nonlinear resistivity
at the peak position is found to be a power law. These results agree
qualitatively with experiments. We also show that the peak of
the nonlinear resistivity is related to the onset of the paramagnetic
Meissner effect which occurs at the crossover temperature $T_p$,
which is above the chiral glass transition temperature $T_{cg}$.
{}~\\
{}~
\end{minipage}
}}

\maketitle



\section{Introduction}

The interplay of superconductivity and disorder in granular superconductors
has been of great interest, particularly regarding their magnetic
properties and glassy behavior.\cite{ebner,choi}
Granular superconductors are usually described as 
a random network of superconducting grains  coupled by Josephson
weak links.\cite{ebner,choi,Joseph,Joseph2} In the last years, there
has been a renewed interest in the study of this problem in
high temperature ceramic superconductors (HTCS).
Several experimental groups have found  
a paramagnetic Meissner effect (PME) 
at low magnetic fields.\cite{pme_exp} 
Sigrist and Rice \cite{sigrist} proposed that this 
effect could be a consequence of the
intrinsic unconventional pairing symmetry 
of the HTCS of $d_{x^2-y^2}$ type.\cite{wollman}
Depending
on the relative orientation of the superconducting grains, it is possible
to have  weak links with negative Josephson coupling \cite{sigrist,wollman}. 
These negative weak links, which are called $\pi$-junctions, 
can give rise to the PME according to Ref.~\onlinecite{pme_exp,sigrist}.
In fact, a model $d$-wave 
granular superconductor, consisting on a network
of Josephson junctions with a random concentration of $\pi$-junctions
and including magnetic screening, 
has been able to explain the paramagenetic Meissner effect
observed at low fields.\cite{Dominguez}
Also in this model, a phase transition to a 
{\it chiral glass} has been predicted for zero magnetic
fields.\cite{KawLi,KawLi1,KawLi2,Kawa}
The frustration effect due to the random distribution of $\pi$ junctions
leads to a glass state of quenched-in ``chiralities'', which are
local loop supercurrents circulating over grains and
carrying a half-quantum of flux.
Evidence of this transition has been related to measurements
of the nonlinear ac magnetic susceptibility.\cite{Matsuura1}
Moreover, the random $\pi$-junction model has also been
adequate to explain several dynamical phenomena observed in HTCS
such as the anomalous microwave absorption,\cite{Dominguez1}
the so called compensation effect \cite{Li} and the effect
of applied electric fields in the apparent critical current.\cite{DWJ}

In recent experiments 
Yamao {\em et al.}\cite{Matsuura} have measured
the ac linear resistivity, $\rho_0$, and the nonlinear resistivity, $\rho_2$,
of ceramic superconductor YBa$_2$Cu$_4$O$_8$. $\rho_0$ and $\rho_2$ are
defined  as the first and third coefficient of the expansion
of the voltage $V(t)$ in terms of the external current $I_{ext}(t)$:
\begin{equation}
V \; \; = \; \rho_0 I_{ext} + \rho_2 I_{ext}^3 + ... \; \; .
\end{equation}
When the sample is driven by an ac current
 $I_{ext}(t) = I_0 \sin(\omega t)$, one can relate $\rho_0$
and $\rho_2$ to the first harmonics, $V'_{\omega}$, and third harmonics,
$V'_{3\omega}$, in the following way
\begin{eqnarray}
\rho_0 \; \; = \; \; V'_{\omega}/I_0, \; \; 
\rho_2 \; \; = \; \; -4 V'_{3\omega}/I_0^3 \; , \nonumber\\
V'_{n\omega} \; \; = \; \; \frac{1}{2\pi}
\int_{-\pi}^{\pi} \; V(t) \sin (n\omega t) 
d(\omega t) \; , \; n \; = \; 1, 3 \; \; .
\end{eqnarray}

The key finding of Ref. \onlinecite{Matsuura} 
is that $\rho_0$ does not vanish
even at and below the intergrain ordering temperature $T_{c2}$.
On the other hand, $\rho_2$ has a peak near this temperature, which
was found to be negative. 
In the chiral glass phase the $U(1)$ gauge symmetry 
is not broken and the phase of the condensate 
remains disordered.\cite{KawLi,KawLi1,KawLi2}
The chiral glass phase, therefore, should not be superconducting but
exhibit an Ohmic behavior with a finite resistance.
Based on these theoretical predictions Yamao {\em et al.}\cite{Matsuura}
speculated that their results give further 
support to the existence of the chiral glass phase, in addition to previous
results from magnetic susceptibility measurements.\cite{Matsuura1}   

Another interesting result of Yamao {\em et al.}\cite{Matsuura}
is the power law dependence
of $|V'_{3\omega}(T_p)/I_0)^3|$ (or of  $\rho_2$)
at its maximum position $T_p$
on $I_0$: $|V'_{3\omega}(T_p)/I_0^3| \sim I_0^{-\alpha}$. 
The experimental value of the power law exponent was
$\alpha\approx1.1$.\cite{expon} 

The goal of our paper is two-fold.
First, we try to reproduce the experimental results \cite{Matsuura}
using the model of the Josephson junctions between $d$-wave 
superconducting grains \cite{Dominguez,KawLi} where the screening 
of the external
field by supercurrents is taken into account.
Second, we discuss the question if
the temperature $T_p$ of the nonlinear resistivity peak
and the transition temperature $T_{cg}$
to the chiral glass phase \cite{KawLi1,KawLi2}
are related. 
We calculate the linear and
nonlinear ac resistivity by a Langevin dynamics simulation.
In agreement with the experimental data \cite{Matsuura}
we find that $\rho_0$ remains finite below and at the temperature 
$T_p$ where $\rho _2$ has a peak. Furthermore, the maximum value of 
$|V'_{3\omega}(T_p)/I_0^3|$
is found to scale with $I_0$ with a power law exponent 
$\alpha = 1.1 \pm 0.6$,
which is close to the experimental value.\cite{Matsuura}
However, we find that $T_p$ is above the equilibrium chiral glass
transition temperature $T_{cg}$. 


\section{Model}

We neglect the charging effects of the grains and
consider the following Hamiltonian\cite{Dominguez,KawLi}
\begin{eqnarray}
{\cal H} = - \sum _{<ij>} J_{ij}\cos (\theta _i-\theta _j-A_{ij})+ \nonumber\\
\frac {1}{2{\cal L}} \sum _p (\Phi_p - \Phi_p^{ext})^2, \nonumber\\
\Phi_p \; \; = \; \; \frac{\phi_0}{2\pi} \sum_{<ij>}^{p} A_{ij} \; , \;
A_{ij} \; = \; \frac{2\pi}{\phi_0} \int_{i}^{j} \, \vec{A}(\vec{r})
d\vec{r} \; \; ,
\end{eqnarray}
where $\theta _i$ is the phase of the condensate of the grain
at the $i$-th site of a simple cubic lattice,
$\vec A$ is the fluctuating gauge potential at each link
of the lattice,
$\phi _0$ denotes the flux quantum,
$J_{ij}$ denotes the Josephson coupling
between the $i$-th and $j$-th grains,
${\cal L}$ is the self-inductance of a loop (an elementary plaquette),
while the mutual inductance between different loops
is neglected.
The first sum is taken over all nearest-neighbor pairs and the
second sum is taken over all elementary plaquettes on the lattice.
Fluctuating  variables to be summed over are the phase variables,
$\theta _i$, at each site and the gauge variables, $A_{ij}$, at each
link. $\Phi_p$ is the total magnetic flux threading through the
$p$-th plaquette, whereas $\Phi_p^{ext}$ is the flux due to an
external magnetic  applied along the $z$-direction,
\begin{equation}
\Phi_p^{ext} = \left\{ \begin{array}{ll}
                   HS \; \;  & \mbox{if $p$ is on the $<xy>$ plane}\\
                   0  & \mbox{otherwise} \; \; ,
                        \end{array}
                  \right.
\end{equation}
where $S$ denotes the area of an elementary plaquette.
For the $d$-wave superconductors
we assume $J_{ij}$ to be an independent random variable
taking the values $J$ or $-J$ with equal probability ($\pm J$ or bimodal
distribution), each representing 0 and $\pi$ junctions.


In order to study the dynamical response and transport properties, we
model the current flowing between two grains with the resistively shunted
junction (RSJ) model,\cite{Joseph,Joseph2} which gives:
\begin{equation}
I_{ij}= \frac{2e}{\hbar}J_{ij}\sin{\theta_{ij}}+\frac{\hbar}{2eR}
\frac{d\theta_{ij}}{dt}+\eta_{ij}(t)
\end{equation}
Here we add to the Josephson current the contribution of a dissipative
ohmic current due to an intergrain resistance $R$ and the Langevin noise
current $\eta_{ij}(t)$ which has correlations
\begin{equation}
\langle\eta_{ij}(t)\eta_{i'j'}(t')\rangle=\frac{2kT}{R}\delta_{i,i'}
\delta_{j,j'}\delta(t-t')
\end{equation}
The dynamical variable in this case is the gauge invariant
phase difference $\theta_{ij}=\theta_i-\theta_j-A_{ij}$. 
The total flux through each plaquette $p$ depends on the 
mesh current $C_p$:
\begin{equation}
\Phi_p=\Phi_p^{ext}+{\cal L}C_p
\end{equation} 
The mesh currents $C_p$, the link currents $I_{ij}$ and the external 
current $I_{ext}$ are related through current conservation.
At this point, it is better to redefine notation:
the site of each grain is at position ${\bf n}=(n_x,n_y,n_z)$
(i.e. $i\equiv{\bf n}$); the lattice directions are
$\mu={\hat{\bf x}}, {\hat{\bf y}}, {\hat{\bf z}}$;
the link variables are between sites ${\bf n}$ and ${\bf n}+\mu$
(i.e. link $ij$  $\equiv$ link ${\bf n},\mu$);
and the plaquettes $p$ are defined by the site ${\bf n}$ and
the normal direction $\mu$ (i.e plaquette $p$ $\equiv$ plaquette
${\bf n},\mu$, for example the plaquette ${\bf n}, {\hat{\bf z}}$ is
centered at position ${\bf n}+({\hat{\bf x}}+{\hat{\bf y}})/2$).
The current $I_\mu({\bf n})$ is related to the mesh
currents $C_\nu({\bf n})$ and the external current in the $y$-direction 
as 
\begin{equation}
I_\lambda({\bf n})=\varepsilon_{\lambda\mu\nu}\Delta_{\mu}^{-}C_\nu({\bf n})
          +\delta_{\lambda,y}I_{ext}
\end{equation}	  
where $\varepsilon_{\lambda\mu\nu}$ is  the Levi-Civita tensor,
$\Delta_{\mu}^{-}$ is the backward difference operator,
$\Delta_{\mu}^{-}C_\nu({\bf n})=C_\nu({\bf n})-C_\nu({\bf n}-\mu)$,
and repeated indices are summed. It is easy to verify that
Eq.(8) satisfies current conservation.
The magnetic flux $\Phi_\lambda({\bf n})$ and the gauge invariante phases
$\theta_\nu({\bf n})=\Delta_\nu^{+}\theta({\bf n})-A_\nu({\bf n})$ 
are related as
\begin{equation}
\Phi_\lambda({\bf n})=-\frac{\Phi_0}{2\pi}\varepsilon_{\lambda\mu\nu}
             \Delta_{\mu}^{+}\theta_\nu({\bf n})
\end{equation}
with the forward difference operator $\Delta_{\mu}^{+}\theta_\nu({\bf n})=
\theta_\nu({\bf n}+\mu)-\theta_\nu({\bf n})$.
	     
Then, from equations (5), (6), (8) and (9) we  obtain 
the following  dynamical equations:
\begin{eqnarray}
\frac{\hbar}{2eR}\frac{d\theta_\mu({\bf n})}{dt}&=&
-\frac{2e}{\hbar}J_\mu({\bf n})\sin\theta_\mu({\bf n})
-\delta_{\mu,y}I_{ext}\nonumber\\ 
& &-\frac{\hbar}{2e{\cal L}}\Delta_\nu^{-}\left[
\Delta_\nu^{+}\theta_{\mu}({\bf n})-\Delta_\mu^{+}\theta_\nu({\bf n})\right]\\
& & -\eta_\mu({\bf n},t)\,\nonumber
\end{eqnarray} 
which represent the RSJ dynamics of a three dimensional Josephson junction 
array with magnetic screening.\cite{Joseph2,Dominguez}

We can also obtain these equations from Eq.~(3) if we add to
${\cal H}$ the external current term:   ${\cal H}_T={\cal H}+
\sum_{\bf n}\frac{\hbar}{2e}I_{ext}\theta_y({\bf n})$. Then
an equation of the Langevin form is obtained by taking derivatives
with respect to the gauge invariant phase difference: 
\begin{equation}
\frac{\hbar}{2eR}\frac{d\theta_\mu({\bf n})}{dt}=
-\frac{2e}{\hbar}\frac{\delta{\cal H}_T\;}{\delta\theta_\mu({\bf n})}
-\eta_\mu({\bf n},t)\;,
\end{equation} 
leading to the RSJ dynamical equations of (10).\cite{tdgl}

In what follows we will consider currents normalized by $I_J=2eJ/\hbar$,
time by $\tau=\phi_0/2\pi I_JR$, voltages by $RI_J$, inductance by
$\phi_0/2\pi I_J$ and temperature by $J/k_B$.	     
We consider open boundary conditions for magnetic fields and currents
in the same way as defined in Ref.\onlinecite{Joseph2,Dominguez}.

\section{Results}

The system of differential equations (10) is integrated numerically by a
second order Runge-Kutta algorithm.
We consider the system size $L=8$ and the self-inductance
${\cal L}=1$. Depending on values of $I_0$ and $\omega$ the number of
samples used for the disorder-averaging ranges between 15 and 40. The 
number of integration steps is chosen to be $10^5 - 5\times 10^5$.

\begin{figure}
\epsfxsize=3.2in
\centerline{\epsffile{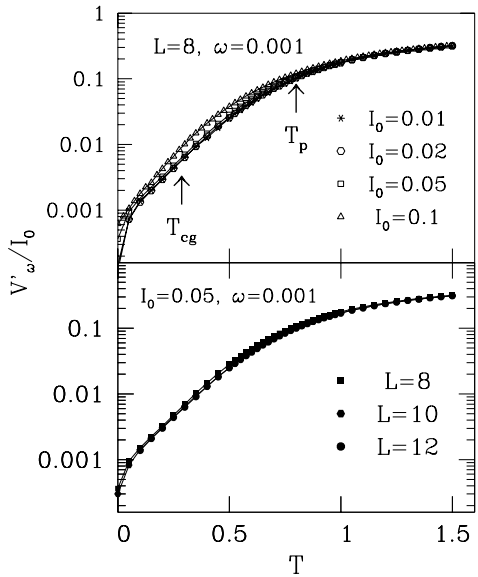}}
\vspace{0.2in}
\caption{a) Upper panel: the temperature dependence of $V'_{\omega}/I_0$ for
$L=8, {\cal L}=1$ and $\omega =0.001$.
The open triangles, squares and hexagons correspond to
$I_0 = 0.1, 0.05, 0.02$ and 0.01. The arrows correspond to $T_p=0.8$
and $T_{cg}=0.286$, respectively.
b) Lower panel: the size dependence of $V'_{\omega}/I_0$ for
$I_0=0.05, {\cal L}=1$ and $\omega =0.001$.
The number of time steps is equal to $10^5$.
The results are averaged over
15 - 40  samples.}
\end{figure}
The temperature dependence of the linear resistivity 
$\rho_0=V'_{\omega}/I_0$ for different values of $I_0$
is shown in Fig. 1 (upper panel). 
At low temperatures we observe a
weak dependence on $I_0$, but for currents small enough $\rho_0$ becomes
independent of current. From the lower panel of Figure 1 it is 
clear that the $V'_{\omega}/I_0$ becomes size-independent for $L>10$.
Thus, the linear resistivity is
nonzero for all temperatures $T>0$ in the thermodynamic limit. 
This is in good agreement with the
result that the U(1) symmetry is not broken in the chiral glass
state,\cite{KawLi,KawLi1,KawLi2} and therefore there is no superconductivity 
for any finite $T$.
We note that a similar result was obtained for the vortex glass state
when the magnetic screening is taken into account.\cite{Bokil}
 
In Fig. 2 we analyze the non-linear resistivity 
$\rho_2=-4V'_{3\omega}(T)/I_0^3$.
We find that it has a negative maximum at a temperature $T_p$.
This characteristic maximum 
depends on $I_0$, but we can fit its position in
temperature at $T_p = 0.8 \pm 0.05$
for all values of $I_0$ presented in Fig. 2.
The arrow in Fig.~1 also indicates the position of  the
temperature $T_p$.  We see that for $T \gg T_p$ the linear resistivity 
$\rho_0$ is independent of current for a large range of currents $I_0$.
On the other hand, below $T_p$ the resistivity $\rho_0$ is current
dependent for an intermediate range of $I_0$ and only for very low 
currents $\rho_0$ becomes current independent.

We identify 
$T_p$  to correspond to the intergrain ordering transition temperature
above which the thermoremanent magnetization disappears in the
experiment of Ref.\onlinecite{Matsuura}.
In order to verify this, we study in Fig.3 the magnetization at
a finite magnetic field $f=HS/\phi_0=0.1$. We show both the zero-field cooling
(ZFC) and field cooling (FC) curves.\cite{Dominguez}
We can see that $T_p$ is the temperature where there is an
onset of positive magnetization, i.e. the paramagnetic  Meissner effect
starts to be observed.
On the other hand, the irreversibility point occurs at temperatures lower
than $T_p$, 
and its position is dependent on the heating or cooling rate.
It should also be noted that above $T_p$ the real part of the linear
magnetic susceptibility vanishes 
(see Fig. 18 from Ref. \onlinecite{KawLi}). 

\begin{figure}
\epsfxsize=3.2in
\centerline{\epsffile{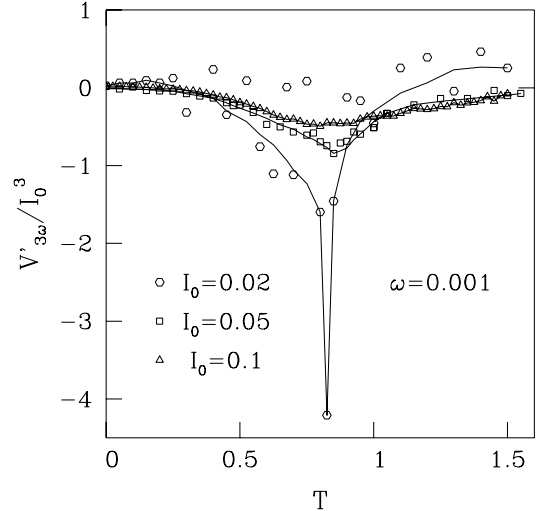}}
\caption{The same as in Fig. 1 but for $|V'_{3\omega}(T_p)/I_0^3|$.} 
\end{figure}

The results presented in Fig. 1, 2 and 3 are in good agreement 
with the experimental data.\cite{Matsuura} 
From this point of view our
findings and the experimental results \cite{Matsuura} may seem
compatible with the chiral-glass picture.\cite{KawLi2} However, 
$T_p$ is remarkably higher than the chiral
glass temperature $T_{cg}$ obtained previously 
(for ${\cal L}=1, T_{cg}=0.286$, see Ref. \onlinecite{KawLi1}). 
Then we conclude that the peak
of $\rho_2$ has no relation to the chiral glass phase transition.
Thus, $T_p$ just separates the normal state phase from a ``chiral paramagnet"
where there are local chiral magnetic moments. These local moments can
be polarized under an external magnetic field, an therefore
one can observe the paramagnetic Meissner effect under
a low external field below $T_p$.  At a lower temperature, collective
phenomena due to the interactions among the chiral moments will start
to be important, leading to the transition to the chiral glass state.
This last transition should show in the nonlinear chiral glass
susceptibility wich should diverge at $T_{cg}$,\cite{KawLi1,KawLi2}. 
The chiral glass transition
may also be reflected  in the irreversibility point
in the FC and ZFC magnetizations. 
Although 
our model is different from the corresponding gauge glass model\cite{Bokil}
one can expect that here the screening spoils any glassy phase except the
chiral glass. The linear resistivity is, therefore, nonzero for 
finite temperatures.

\begin{figure}
\epsfxsize=3.2in
\centerline{\epsffile{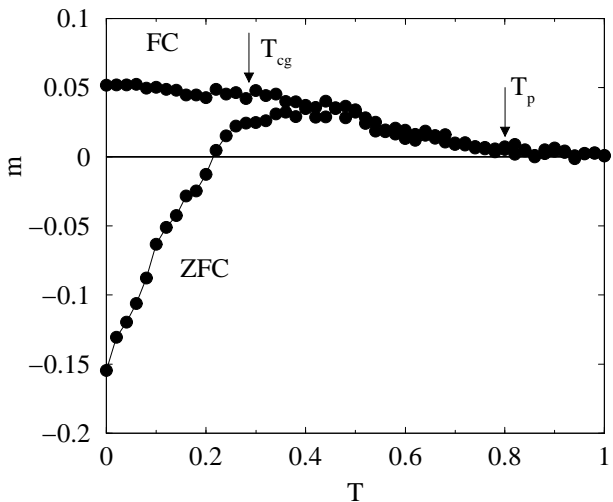}}
\caption{The temperature dependence of the magnetization $m$ in
FC and ZFC regimes for the $d$-wave superconductors. $L=8$ and ${\cal L}=1$.
The results are averaged over 25 samples.}
\end{figure}

Our calculation of the nonlinear ac resistivity $\rho_2$ is a non-equilibrium
calculation at a finite frequency $\omega$ and finite ac current amplitude
$I_0$. Therefore, one should be concerned on the finite $\omega$ and
finite $I_0$ effects. In particular, one may ask if  it is possible that the
tempereture $T_p$ of the peak in $\rho_2$
will tend to $T_{cg}$ in the limit  $\omega\rightarrow0$, $I_0\rightarrow0$.
We have carefully studied this possibility.
Fig. 4 shows the temperature dependence of $\rho_2$ for various
values of $\omega$ and $I_0=0.1$. From Fig. 2 and 4 it is clear
that the position of $T_p$ depends on $I_0$ and $\omega$ very weakly.
It is, therefore, unlikely
that $T_p$ tends to $T_{cg}$ as $\omega\rightarrow0$ and $I_0\rightarrow0$.

In accordance with the experiments of Yamao
{\em et al.}, \cite{Matsuura} the negative maximum of
$V'_{3\omega}(T_p)/I^3$ is shown up.
Furthermore, the height of peaks
of $|V'_{3\omega}(T_p)/I^3|$ increases with the decrease of $\omega$
and saturates at small frequencies (see Fig. 4).
Such tendency was also observed experimentally. \cite{Matsuura}

In order to get more insight on the nature of $T_p$ we have calculated
the specific heat, $C_v$, which is defined as porportional to
the energy fluctuations, $C_v=\langle(\delta E)^2\rangle/k_BT^2$.
The results are shown in Fig. 5. There is a broad peak in $C_v$
located at $T_p$ and well above $T_{cg}$.
Similar to the spin glass case where the peak of
specific heat is positioned higher than the critical temperature
to the glass phase ,\cite{Binder}
we conclude that $T_p$ does not correspond to a phase transition
to a long-range ordered phase.

\begin{figure}
\epsfxsize=3.2in
\centerline{\epsffile{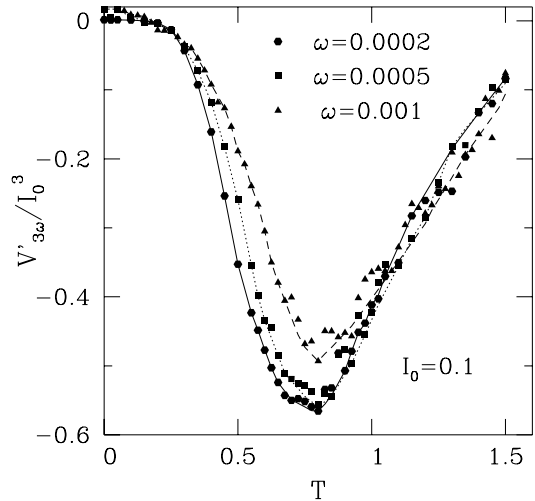}}
\caption{The temperature dependence of $V'_{3\omega}(T)/I_0^3$.
The solid triangles, squares and hexagons correspond
to $\omega = 0.001, 0.0005$ and $0.0002$, respectively. $L=8, {\cal L}=1$
and $I_0=0.1$. The results are averaged over 15 samples.}
\end{figure}

\begin{figure}
\epsfxsize=3.2in
\centerline{\epsffile{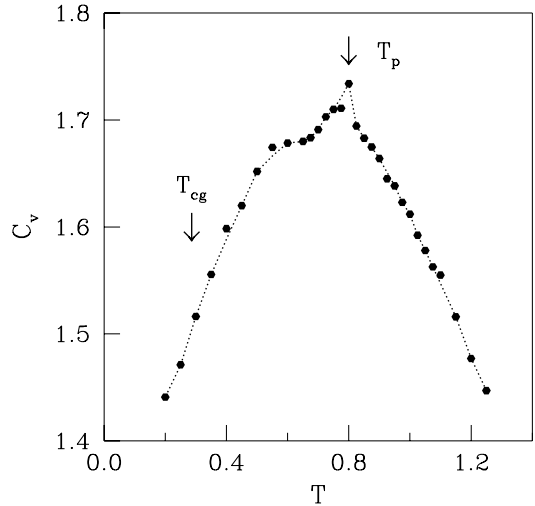}}
\caption{The temperature dependence of $C_v$ obtained by Monte Carlo
simulations for $L=8$ and ${\cal L}=1$.
The results are averaged over 20 samples. The error bars are smaller than
symbol sizes.}
\end{figure}

A more convincing conclusion about the nature of the peak in the nonlinear
susceptibility should be obtained from a finite size  analysis.
Fig. 6 shows the dependence of $max|V'_{3\omega}/I_0^3|$ on the system size
$L$ for $I_0=0.05$ and $\omega=0.001$. Clearly, the height of the peak
does not diverge as $L \rightarrow \infty $. In other words the peak
in the nonlinear resistivity does not corresponds to a phase transition
in the thermodynamic limit.

\begin{figure}
\epsfxsize=3.2in
\centerline{\epsffile{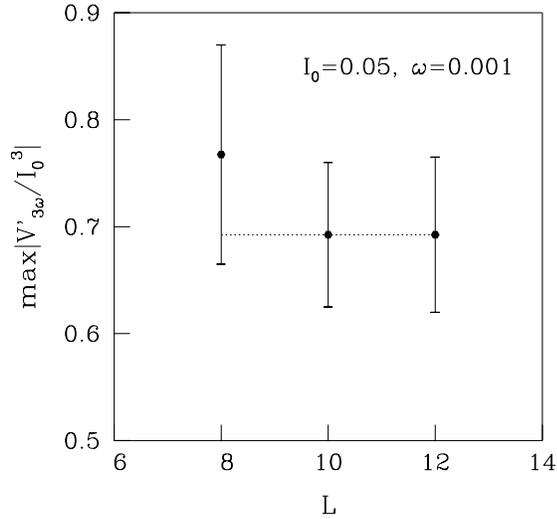}}
\caption{The dependence of the maximal values of $|V'_{3\omega}(T_p)/I_0^3|$
on the system size $L$.
$I_0=0.005, {\cal L}=1$
and $\omega =0.001$. The results are averaged over 15 - 40 samples.}
\end{figure}

\begin{figure}
\epsfxsize=3.2in
\centerline{\epsffile{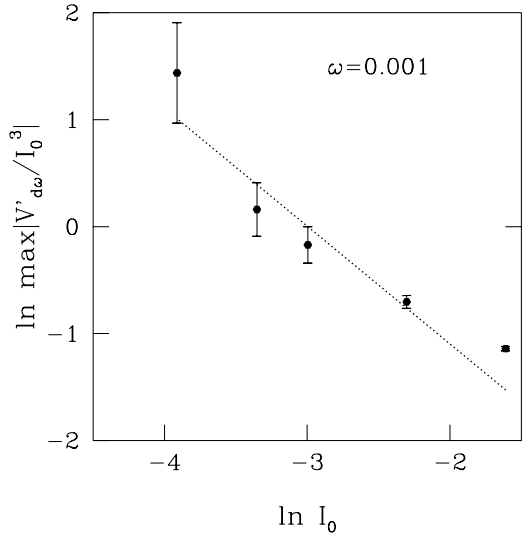}}
\vspace{0.15in}
\caption{The dependence of the maximal values of $|V'_{3\omega}(T_p)/I_0^3|$
on $I_0$.
$L=8, {\cal L}=1$
and $\omega =0.001$. The results are averaged over 15 - 40 samples.}
\end{figure}

Fig.~7 shows  the log-log plot for
the  dependence of the maximal values of $|V'_{3\omega}(T_p)/I_0^3|$
on $I_0$ for a fixed frequency $\omega=0.001$.
One can fit $\max|V'_{3\omega}(T_p)/I_0^3| \sim I_0^{\alpha}$
with $\alpha = 1.1 \pm 0.6$ giving more weight to small values of $I_0$.
So within the error bars our estimate of $\alpha$  agrees
with that obtained by
the experiments. \cite{Matsuura,expon}

\section{Discussion}

In Ref. \onlinecite{Matsuura}
it was argued that the peak of the nonlinear resistivity
was a signal of the transition to the chiral glass state. 
The value of $T_p$ obtained in our simulations is, however,
considerably higher than the 
chiral glass transition
temperature, $T_{cg}$. We conclude that the peak of
$\rho_2$ is not related to the transition to the chiral glass.
$T_p$ is found to coincide with the point 
for the onset of the paramagnetic Meissner effect, where
the magnetization becomes positive.
In this respect, our result
agrees with the experimental result.\cite{Matsuura} 
We interpret $T_p$ as the crossover temperature from the normal state phase to
a ``chiral paramagnet" in which there are local chiral magnetic moments
induced by the $\pi$-junctions. 
As the temperature is lowered down the system would have a phase transition
from the chiral paramagnetic phase to the chiral glass state. 
At this critical point $\rho_2$ does not show any particular feature.
Furthermore, we found that the linear resistivity is always finite at $T>0$ 
due to screening effects, and therefore there is no superconductivity
in the random $\pi$ junction model.

In conclusion, the experimental results of Yamao {\em et al.} \cite{Matsuura}
can be reproduced by the XY-like model for the $d$-wave superconductors.
Contrary to the speculation of Ref. \onlinecite{Matsuura} we expect that
$T_p$ does not correspond to the chiral glass transition.

\acknowledgments

One of us (MSL) thanks M. Cieplak,
H. Kawamura, A. Majhofer, T. Nattermann and M. Sigrist for useful discussions.
Financial support from the Polish agency KBN
(Grant number 2P03B-146-18) is acknowledged. D. D. acknowledges
financial support from Fundaci\'{o}n Antorchas 
(Proy. A-13532/1-96), ANPCyT (PICT-03-00121-02151), Conicet and CNEA
(Argentina). We also acknowledege financial support from
the International Centre for Theoretical Physics.

\par


\begin{references}
\bibitem{expon} We have extracted the experimental value of $\alpha\approx 1.1$
from Fig.~7 of Ref.~\protect\onlinecite{Matsuura}.
\end{references}
\end{document}